\documentclass[%
 reprint,
 superscriptaddress,
 amsmath,amssymb,
 aps,
]{revtex4-2}

\usepackage{graphicx}
\usepackage{dcolumn}
\usepackage{bm}
\usepackage{hyperref}
\usepackage{cleveref}
\usepackage{enumerate}
\usepackage{relsize}
\usepackage[mathlines]{lineno}

\allowdisplaybreaks

\usepackage{custom}
\usepackage{tabularx}

\newcommand{\hmu}{\hat{\mu}}

\newcommand{\MeV}{\, \rm{MeV}}

\begin{document}

\preprint{APS/123-QED}

\title{
A new 4D lattice QCD equation of state: extended density coverage from a generalized $T^\prime$-expansion
}

\author{Ahmed Abuali}
\affiliation{
 Department of Physics, University of Houston, Houston, TX 77204, USA
}

\author{Szabolcs Bors\'anyi}
\affiliation{Department of Physics, Wuppertal University, Gaussstr.  20, D-42119, Wuppertal, Germany}

\author{Zolt\'an Fodor}
\affiliation{Pennsylvania State University, Department of Physics, State College, PA 16801, USA}
\affiliation{Pennsylvania State University, Institute for Computational and Data Sciences, State College, PA 16801, USA}
\affiliation{Department of Physics, Wuppertal University, Gaussstr.  20, D-42119, Wuppertal, Germany}
\affiliation{Institute  for Theoretical Physics, ELTE E\"otv\"os Lor\' and University, P\'azm\'any P. s\'et\'any 1/A, H-1117 Budapest, Hungary}
\affiliation{J\"ulich Supercomputing Centre, Forschungszentrum J\"ulich, D-52425 J\"ulich, Germany}

\author{Johannes Jahan} \email[Corresponding author: ]{jahan.johannes@gmail.com}
\affiliation{
 Department of Physics, University of Houston, Houston, TX 77204, USA
}

\author{Micheal Kahangirwe}
\affiliation{
 Department of Physics, University of Houston, Houston, TX 77204, USA
}

\author{Paolo Parotto}
\affiliation{Dipartimento di Fisica, Universit\`a di Torino and INFN Torino, Via P. Giuria 1, I-10125 Torino, Italy}

\author{Attila P\'asztor}
\affiliation{Institute  for Theoretical Physics, ELTE E\"otv\"os Lor\' and University, P\'azm\'any P. s\'et\'any 1/A, H-1117 Budapest, Hungary}

\author{Claudia Ratti}
\affiliation{
 Department of Physics, University of Houston, Houston, TX 77204, USA
}

\author{Hitansh Shah}
\affiliation{
 Department of Physics, University of Houston, Houston, TX 77204, USA
}

\author{Seth A. Trabulsi}
\affiliation{
 Department of Physics \& Astronomy, Rice University, 6100 Main St., Houston, TX 77005, USA
}

\date{\today}

\begin{abstract}
We present a new equation of state for QCD in which the temperature $T$ and the three chemical potentials for baryon number $\mu_B$, electric charge $\mu_Q$ and strangeness $\mu_S$ can be varied independently. This result is based on a generalization of the $T'$-expansion scheme, thanks to which the diagonal $\mu_B$ extrapolation was pushed up to a baryo-chemical potential $\mu_B/T \sim 3.5$ for the first time. This considerably extended the coverage of the Taylor expansion, limited to $\mu_B/T < 2.5-3$. As a consequence, we are able to offer a substantially larger coverage of the four-dimensional QCD phase diagram as well, compared to previously available Taylor expansion results. Our results are based on new continuum estimated lattice results on the full set of second and fourth order fluctuations.
\end{abstract}


\maketitle

\section{\label{sec:level1}INTRODUCTION\protect\\ }

The description of the phases of strongly interacting matter under different conditions is a major task of nuclear physics, that encompasses a vast amount of systems ranging from the early universe to the interior of compact objects like neutron stars. 
It is customary to summarize this knowledge into a phase diagram in the temperature $T$ and baryon chemical potential $\mu_B$ (or baryon density $n_B$) plane. Explorations of the phase diagram of QCD have been carried out extensively by both theory and experiment in the past decades. It is known that a low-temperature/low-density hadron gas phase is separated from a high-temperature quark gluon plasma (QGP) phase. At exactly zero density, the transition was shown to be an analytic crossover~\cite{Aoki:2006we} at around $T \simeq 155-160$~MeV \cite{Aoki:2006br,Aoki:2009sc,Borsanyi:2010bp,Bazavov:2011nk,Bazavov:2018mes,Borsanyi:2020fev}. 
At small temperatures and densities of the order of nuclear saturation density, neutron star matter is expected to exist, although its characterization is still ongoing. At even larger densities, a rich phase structure is expected for QCD matter~\cite{Alford:2007xm}. 

The thermodynamic properties of a system such as strongly interacting matter are encapsulated in its equation of state, which is an essential input for modeling and effectively describing QCD matter.  The equation of state plays a crucial role for hydrodynamic simulations of heavy ion collisions and neutron star mergers, where it is necessary to close the set of conservation equations to be solved. The equation of state of QCD is known with good precision at vanishing density from lattice QCD~\cite{Borsanyi:2013hza,Bazavov:2014pvz,Borsanyi:2025dyp}.
Lattice simulations are the main first-principles method to investigate the thermodynamics of QCD. They amount to numerically calculate the path integral in the euclidean formulation by means of Monte Carlo methods. Many of the established features of the phase diagram have been determined via lattice simulations, like the crossover nature of the  transition, and its pseudo-critical temperature. Because of the fermion sign problem, direct lattice simulations at non-zero density are extremely costly, and are currently limited to small volumes \cite{Alexandru:2005ix,Li:2010qf,Borsanyi:2021hbk,Borsanyi:2022soo}. 
Most results on finite-density thermodynamics rely on extrapolations, namely Taylor expansion or analytic continuation from imaginary chemical potentials \cite{deForcrand:2002ci,DElia:2002tig,Bonati:2015bha,DElia:2016jqh,Gunther:2016vcp,Bonati:2018nut,Borsanyi:2018grb,Bellwied:2015rza,Bollweg:2020yum,Borsanyi:2020fev,Bollweg:2022rps,Ding:2024sux}.
Both currently allow for an exploration of the phase diagram at small to intermediate densities. For the equation of state, Taylor expansions around $\mu_B=0$ up to N$^3$LO allow to describe the regime $\mu_B/T\leq 2.5-3$ \cite{Bollweg:2022fqq}, though at higher densities they typically break down due to the presence of unphysical behavior in some thermodynamic quantities. This is likely due to the fact that the extrapolation is carried out at constant temperature, and is thus forced to cross the transition line at some non-zero $\mu_B$. 
Recently, an alternative expansion has been developed~\cite{Borsanyi:2021sxv,Borsanyi:2022qlh}, that implements the extrapolation along lines of constant density, thus avoiding crossing the transition. Such alternative scheme, also dubbed $T^\prime$-expansion scheme (TExS), allows for extrapolations up to $\mu_B/T\leq 3.5$, with small uncertainties and without unphysical behavior. It also shows better convergence properties than Taylor expansions, due to a large separation between the LO and NLO expansion coefficients. 
Recently, in Ref.~\cite{Wen:2024hbz,Kahangirwe:2024xyl} this scheme was compared to Taylor expansions and Padé approximants in a number of models of QCD matter, showing superior convergence properties around and above the transition temperature
at finite $\mu_B$ and along the strangeness neutral trajectory, though not in the hadronic phase. A comparison with direct lattice QCD results at finite chemical potential from reweighting methods was presented in Ref.~\cite{Borsanyi:2022soo}, showing substantial agreement. 

Although it is customary to depict the phase diagram in the $T-\mu_B$ plane, it is actually a four-dimensional space, where three chemical potentials can be varied, associated to the 
conserved charges of QCD: baryon number $B$, electric charge $Q$ and strangeness $S$.
Most commonly, results at finite density are extrapolated either along the $\mu_Q = \mu_S = 0$ direction, or along the strangeness neutral line, defined by $n_S = 0$ and $n_Q = 0.4 \, n_B$, to reproduce the experimental setting of heavy ion collision systems.

However, for realistic simulations of QCD matter, a comprehensive description of the full 4D space $T,\mu_B,\mu_Q,\mu_S$ is necessary. It was shown in Ref.~\cite{Plumberg:2024leb} that even in LHC settings, where all densities vanish on average, local fluctuations lead to a large range of chemical potentials to be sampled by individual fluid cells.

In order to address the need for a 4D equation of state, in Ref.~\cite{Noronha-Hostler:2019ayj} a Taylor expansion of the QCD pressure was constructed, based on lattice QCD results at $N_\tau=12$ for the susceptibilities~\cite{Borsanyi:2018grb}, and in the continuum for the zero-density equation of state~\cite{Borsanyi:2013hza}:
\begin{equation}
    \frac{p(T,\hat{\mu}_B,\hat{\mu}_Q,\hat{\mu}_S)}{T^4} = \sum_{i,j,k} \frac{1}{i!j!k!}\chi_{ijk}^{\text{BQS}}(T) \hat{\mu}_B^i \hat{\mu}_Q^j \hat{\mu}_S^k \, \, ,
\label{eq:TaylorExpansion}
\end{equation}
where $\hat{\mu}_i=\frac{\mu_i} {T}$ ($i=\{B,Q,S\}$), and the coefficients are the susceptibilities:
\begin{equation}
    \chi_{ijk}^{\text{BQS}}(T) = \frac{\partial^{i+j+k} (p/T^4)}{\partial \hat{\mu}_B^i\hat{\mu}_Q^j\hat{\mu}_S^k}\Bigg|_{\hat{\mu}_B,\hat{\mu}_Q,\hat{\mu}_S=0} \, \, .
    \label{eq:susceptibilities}
\end{equation}

The lattice susceptibilities in Eq. \eqref{eq:susceptibilities} were merged with results from the hadron resonance gas (HRG) model~\cite{Dashen:1969ep} to cover the low-temperature range relevant for hydrodynamic simulations. At high temperatures, the approach to the Stefan-Boltzmann (SB) limit was imposed on each susceptibility. The resulting equation of state, expanded to order ${\cal O} (\mu^4)$ (i.e. $i+j+k\leq 4$), was made available in the range $T = 30 - 800 $ MeV, $\mu_k < 450 \, \text{MeV}$. A similar construction was presented in Refs.~\cite{Monnai:2019hkn,Monnai:2024pvy}, based on susceptibilities up to order ${\cal O} (\mu^4)$ and some of order ${\cal O} (\mu^6)$, from Refs.~\cite{HotQCD:2012fhj,Ding:2015fca,Bazavov:2017dus}, and the zero-density equation of state from Ref.~\cite{HotQCD:2014kol}. In both constructions, the 4D equation of state showcases the same limitations in the range reliably covered by the expansion.

In this work, we construct a new 4D equation of state for QCD covering a broader range of densities accessible in high-energy heavy-ion collisions, by generalizing the expansion scheme of Ref.~\cite{Borsanyi:2021sxv} to the case with three independent chemical potentials. 
We base our construction on continuum estimated susceptibilities obtained with the 4stout action on lattices with $N_\tau = 10, 12, 16, 20, 24$ time slices, with an aspect ratio $LT=4$ except for the  $64^3 \times 24$ lattice. This is the first equation of state of QCD in the 3D space of chemical potentials to be based entirely on continuum estimated susceptibilities and providing an estimate of the related uncertainties. Working with a large volume, we can expect finite size effects to be small~\cite{Borsanyi:2025lim}. Details on the action used to obtain the susceptibilities and its parametrization can be found in Ref.~\cite{Bellwied:2015lba}.

We go over the basics of the $T^\prime$-expansion scheme and formulate its generalization to any direction in the chemical potential space in Section~\ref{sec:altexp}. 
In Section~\ref{sec:susc} we present our results for the continuum estimated susceptibilities of order 2 and 4. 
We show results for the different thermodynamic quantities in Section~\ref{sec:results}, together with a discussion on the limits of applicability of our expansion, before presenting our conclusions in Section~\ref{sec:conclusions}.

\section{\label{sec:altexp}$T^\prime-$EXPANSION SCHEME IN 4D\protect\\ }

\begin{figure*}
    \centering
    \includegraphics[width=0.9\textwidth]{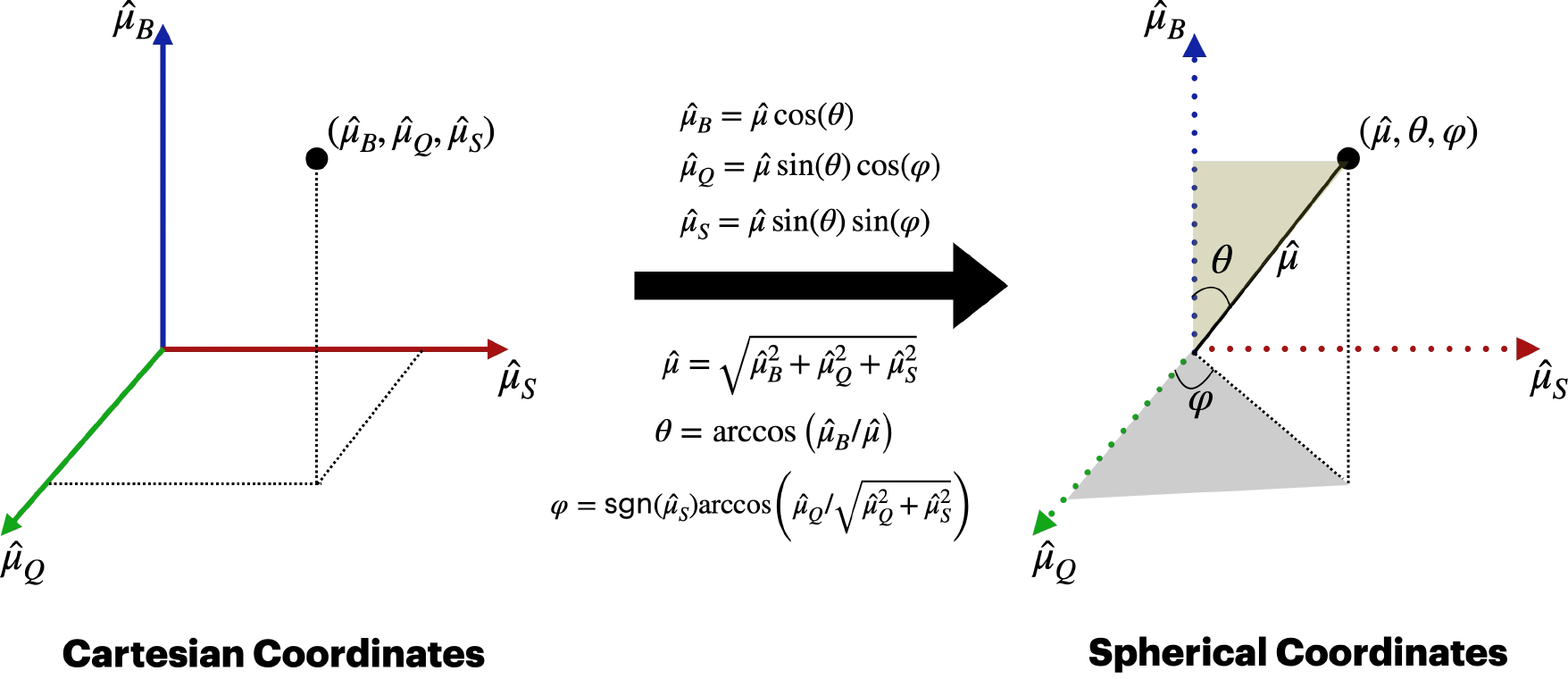}
    \caption{The 3D space of chemical potentials in cartesian ($\hmu_B,\hmu_Q,\hmu_S$) (left) and spherical ($\hmu,\theta,\varphi$) (right) coordinates. With the latter, we can perform a 1D extrapolation in any direction.}
    \label{fig:mapping}
\end{figure*}

The idea behind the $T^\prime-$expansion scheme introduced in Refs.~\cite{Borsanyi:2021sxv,Borsanyi:2022qlh} is to take advantage of the fact that the dependence of certain fluctuation observables on the chemical potential $\mu_B$  largely amounts to a $\mu_B$-dependent temperature shift. In particular, it was noticed that the normalized baryon density $\chi_1^B/\hmu_B$ at imaginary chemical potential resembles the second baryon cumulant at $\mu_B=0$, with a simple redefinition of the temperature.
The expansion scheme is defined rigorously as:
\begin{equation}
    \frac{n_B(T,\hat{\mu}_B)}{\hat{\mu}_B} = \chi_2^B(T',0) \, \, ,
    \label{eq:1DAltExS}
\end{equation}
with the re-defined temperature
\begin{equation}
    T'(T,\hmu_B) = T\left(1+ \kappa_2^B (T) \hmu_B^2 + \ldots \right).
	\label{eq:kappaN}
\end{equation}

In Ref.~\cite{Borsanyi:2022qlh}, the scheme was applied to the strangeness neutral setting, where non-zero values of $\mu_Q$ and $\mu_S$ are introduced to ensure $n_S=0$ and $n_Q = 0.4 n_B$. These conditions define a $T$-dependent trajectory in the 3D space of chemical potentials for the expansion. 
Additionally, Eq.~\eqref{eq:1DAltExS} was modified by considering on both sides of the equation quantities normalized by their own infinite temperature limits:
\begin{equation}
    \frac{F(T,\hat{\mu}_B)}{\overline{F}(\hmu_B)} = \frac{F(T_F^\prime,0)}{\overline{F}(0)} \, \, ,
    \label{eq:2D-TExS_SB}
\end{equation}
where the SB limits are indicated by barred quantities, and their argument is the chemical potential. 
The observable-dependent effective temperature $T^\prime_F$ reads:
\begin{equation}
    T^\prime_F(T,\hmu_B) = T \left(1+ \lambda_{2,F} (T) \hmu_B^2 + \ldots \right).
    \label{eq:2D-Tprime}
\end{equation}

Note that the $\lambda_n (T)$ coefficients in Eq.~\eqref{eq:2D-Tprime} are different from the $\kappa_n(T)$ coefficients in Eq.~\eqref{eq:kappaN}. In fact, the expansions defined by Eqs.~\eqref{eq:1DAltExS} and~\eqref{eq:2D-TExS_SB} are essentially two different physics-motivated rearrangements of the Taylor expansion, and the coefficients $\kappa_n (T)$, $\lambda_n (T)$ are in direct correspondence with the Taylor coefficients $\chi_n(T)$. At LO one has \cite{Borsanyi:2021sxv,Borsanyi:2022qlh} (we drop the superscript $B$, as this applies in 
general):
\begin{align}
    \kappa_2 (T) &= \frac{1}{6 T \chi_2^{ \prime} (T)} \chi_4 (T) \, \, , \\
    \lambda_2 (T) &= \frac{1}{6 T \chi_2^{\prime} (T)} \left( \chi_4 (T) - \frac{\overline{\chi_4}}{\overline{\chi_2}} \chi_2(T)\right) \, \, , 
    \label{eq:chistokappalambda}
\end{align}
where $\chi_2^{\prime}$ is the $T$ derivative of $\chi_2$ and the barred quantities again indicate the SB limit values. By imposing the equality on quantities normalized by their SB limit, the high temperature behavior is automatically encoded in the SB-corrected expansion defined by Eq.~\eqref{eq:2D-TExS_SB}. 
It was shown that, in the case of baryon density, $\kappa_2^B  (T)$ and $\lambda_2^B(T)$ start to differ substantially around the transition temperature, where the second term in $\lambda_2^B(T)$ becomes relevant.

\begin{figure*}[!ht]
    \includegraphics[width=\textwidth]{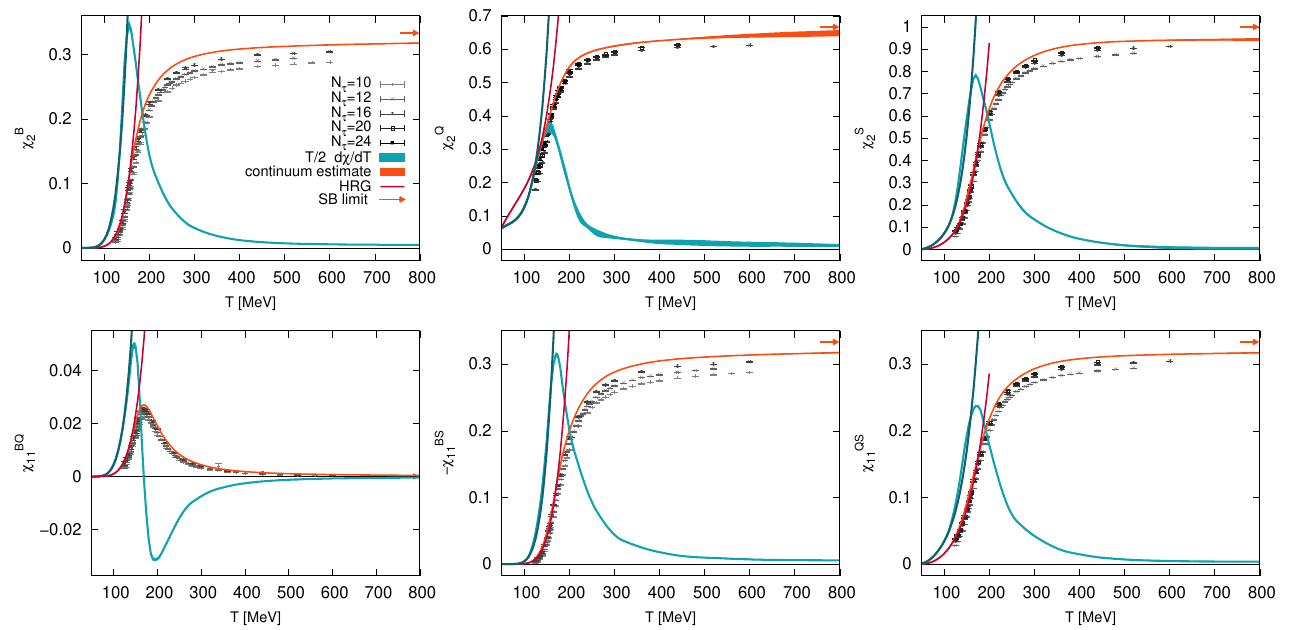}
    \caption{Continuum estimate of all second order susceptibilities (orange), along with the corresponding $T$-derivatives (blue). The latter are multiplied by a factor $1/2$ for ease of comparison. The Stefan-Boltzmann limits of the susceptibilities are indicated by the blue arrows. The gray points show the finite-$N_\tau$ lattice data used to construct the estimate. The solid lines show the ideal HRG model predictions.
    } \label{fig:chi_o2}
\end{figure*}

We define here a generalization of Eq.~\eqref{eq:1DAltExS}, whereby we perform the expansion in an arbitrary direction in the 3D space of chemical potentials. We use spherical coordinates as shown in Fig.~\ref{fig:mapping}:
\begin{align} \nonumber     \label{eq:mustoangles}
    \hat{\mu}_B &= \hat{\mu} \, c_{\theta} \, \, ,\\
    \hat{\mu}_Q &= \hat{\mu} \, s_{\theta} c_{\varphi} \, \, , \\ \nonumber
    \hat{\mu}_S &= \hat{\mu} \, s_{\theta} s_{\varphi} \, \, ,
\end{align}
where $s_{\gamma} = \sin\gamma$ and $c_{\gamma} = \cos\gamma$, and:
\begin{align} \nonumber     \label{eq:anglestomus}
    \hat{\mu} &= \sqrt{\hat{\mu}_B^2 + \hat{\mu}_Q^2 + \hat{\mu}_S^2} \, , \\
    \theta  &= \arccos\left(\hat{\mu}_B/\hat{\mu}\right) \, , \\ 
    \varphi &= \text{sgn}(\hmu_S) \times \arccos\!\left(\hat{\mu}_Q/\sqrt{\hat{\mu}_Q^2 + \hat{\mu}_S^2}\right) \, . \nonumber
\end{align}

With this prescription, given any set of $(\hmu_B,\hmu_Q,\hmu_S)$, the angles $\theta$, $\varphi$ 
are fixed and we are faced with a 1D extrapolation in $\hmu$ along the specified direction. In analogy with Eq.~\eqref{eq:2D-TExS_SB}, we can define the generalized expansion:
\begin{align}
    X_1^{\theta,\varphi}(T, \hat{\mu}) =
    \frac{\overline{X}_1^{\theta,\varphi}( \hat{\mu})}{\overline{X}_2^{\theta,\varphi}(0)}
    X_2^{\theta,\varphi}(T^{\, \prime \, \theta,\varphi}(T,\hat{\mu}), 0)
    \; 
    \label{eq:4D-TExS}
\end{align}
where $X_1^{\theta,\varphi}$ and $X_2^{\theta,\varphi}$ are the generalized density and second order susceptibility along the direction specified by the angles $(\theta,\varphi)$. As before, barred quantities refer to SB limits, with their argument being the chemical potential. 
The effective temperature is defined, in analogy with Eq.~\eqref{eq:2D-Tprime}, as:
\begin{equation}
    T^{\, \prime \, \theta,\varphi} (T,\hmu) = T \left( 1 + \lambda_{2}^{\theta,\varphi} (T) \hmu^2 + \ldots \right) \, ,
    \label{eq:4D-Tprime}
\end{equation}
where the expansion coefficient $\lambda_2^{\theta,\varphi}(T)$ now reads:
\begin{align}
\lambda_{2}^{\theta,\varphi} (T) &= \frac{1}{6 T (X_2^{\theta,\varphi} (T))^{\prime}} \! \left( \! X_4^{\theta,\varphi}  (T) - \frac{\overline{X_4}^{\theta,\varphi}(0)}{\overline{X_2}^{\theta,\varphi}(0)} X_2^{\theta,\varphi}(T) \! \right) . 
\label{eq:generalizedlambda}
\end{align}
There, $X_4^{\theta,\varphi}(T)$ is the generalized fourth order susceptibility along the direction specified by $(\theta, \varphi)$. Note that, when we refer to quantities at $\hmu=0$, we drop the $\hat{\mu}$ dependence to alleviate the notation. 

The quantities $X_n^{\theta,\varphi}(T)$ generalize the susceptibilities $\chi_n(T)$, as they are derivatives of the QCD pressure with respect to the direction-dependent generalized chemical potential $\hmu$:
\begin{equation}
    X_n^{\theta,\varphi} (T) = \frac{\partial^n p/T^4}{\partial \hmu^n} \Bigg|_{\hmu=0}^{\theta,\varphi} \, \, ,
\end{equation}
where the derivative is a directional derivative along the direction identified by the angles $({\theta,\varphi})$.
The $X^{\theta,\varphi}_n$ at $\hmu=0$ are related to the susceptibilities $\chi_{ijk}^{BQS}(T)$ by straightforward application of the chain rule:
\begin{eqnarray}
\label{eq:X1}
    &&X_1^{\theta,\varphi}(T) = 
    c_{\theta} \chi_1^B(T) + s_{\theta} c_{\varphi} \chi_1^Q(T) + s_{\theta} s_{\varphi} \chi_1^S(T) \, \, , \\[1em]
%
\label{eq:X2}
    && X_2^{\theta,\varphi}(T) = 
    c_{\theta}^2\chi_2^B(T) + s_{\theta}^2c_{\varphi}^2\chi_2^Q(T) + s_{\theta}^2 s_{\varphi}^2\chi_2^S(T) \\
    \nonumber
    && \quad + 2c_{\theta}s_{\theta}c_{\varphi}\chi_{11}^{BQ}(T)+ 2c_{\theta}s_{\theta}s_{\varphi}\chi_{11}^{BS}(T) 
    + 2s_{\theta}^2c_{\varphi}s_{\varphi}\chi_{11}^{QS}(T) \, \, , 
    \\[1em]
%
\label{eq:X4}
    && X_4^{\theta,\varphi}(T) 
    = c_{\theta}^{4} \chi_{4}^{B}(T) 
    + s_{\theta}^{4} c_{\varphi}^{4} \chi_{4}^{Q}(T) 
    + s_{\theta}^{4} s_{\varphi}^{4} \chi_{4}^{S}(T) \nonumber
     \\ \nonumber && \quad
    + 4 c_{\theta} s_{\theta}^{3} c_{\varphi}^{3} \chi_{13}^{BQ}(T)  
    + 4 c_{\theta} s_{\theta}^{3} s_{\varphi}^{3} \chi_{13}^{BS}(T) 
    + 4 s_{\theta}^{4} c_{\varphi} s_{\varphi}^{3} \chi_{13}^{QS}(T) 
    \\ \nonumber && \quad 
    + 4 c_{\theta}^{3} s_{\theta} c_{\varphi} \chi_{31}^{BQ}(T) 
    + 4 c_{\theta}^{3} s_{\theta} s_{\varphi} \chi_{31}^{BS}(T) 
    + 4 s_{\theta}^{4} c_{\varphi}^{3} s_{\varphi} \chi_{31}^{QS}(T)
    \\ \nonumber && \quad 
    + 6 c_{\theta}^{2} s_{\theta}^{2} c_{\varphi}^{2} \chi_{22}^{BQ}(T) 
    + 6 c_{\theta}^{2} s_{\theta}^{2} s_{\varphi}^{2} \chi_{22}^{BS}(T) 
    + 6 s_{\theta}^{4} c_{\varphi}^{2} s_{\varphi}^{2} \chi_{22}^{QS}(T) 
    \\ \nonumber && \quad 
    + 12 c_{\theta} s_{\theta}^{3} c_{\varphi} s_{\varphi}^{2} \chi_{112}^{BQS}(T) 
    + 12 c_{\theta} s_{\theta}^{3} c_{\varphi}^{2} s_{\varphi} \chi_{121}^{BQS}(T) 
    \\  && \quad
    + 12 c_{\theta}^{2} s_{\theta}^{2} c_{\varphi} s_{\varphi} \chi_{211}^{BQS}(T) \, \, .
\end{eqnarray}

In this work, we carry out the expansion up to $\lambda_2^{\theta,\varphi}(T)$, namely at leading order, which however corresponds to NLO in the Taylor expansion, including susceptibilities up to order four. The procedure can be easily generalized to higher orders, provided that higher order susceptibilities are available. 

In practice, given a set $\hmu_B,\hmu_Q,\hmu_S$ at a certain temperature $T$, we will go 
through the following steps:
\begin{enumerate}[\bf i.]
    \item determine $\hmu,\theta,\varphi$;
    \item construct $\lambda_2^{\theta,\varphi}(T)$ from $X_2^{\theta,\varphi}(T)$, $X_4^{\theta,\varphi}(T)$ using Eq.~\eqref{eq:generalizedlambda};
    \item determine $X_1^{\theta,\varphi}(T,\hmu)$ using Eq.~\eqref{eq:4D-TExS}.
\end{enumerate}

From $X_1^{\theta,\varphi}(T,\hmu)$, together with the zero-density equation of 
state, all thermodynamic quantities can be calculated, as will be detailed in 
Section~\ref{subsec:thermo_finitemu}.

\section{Susceptibilities from the lattice}\label{sec:susc}


\begin{figure*}
    \includegraphics[width=\textwidth]{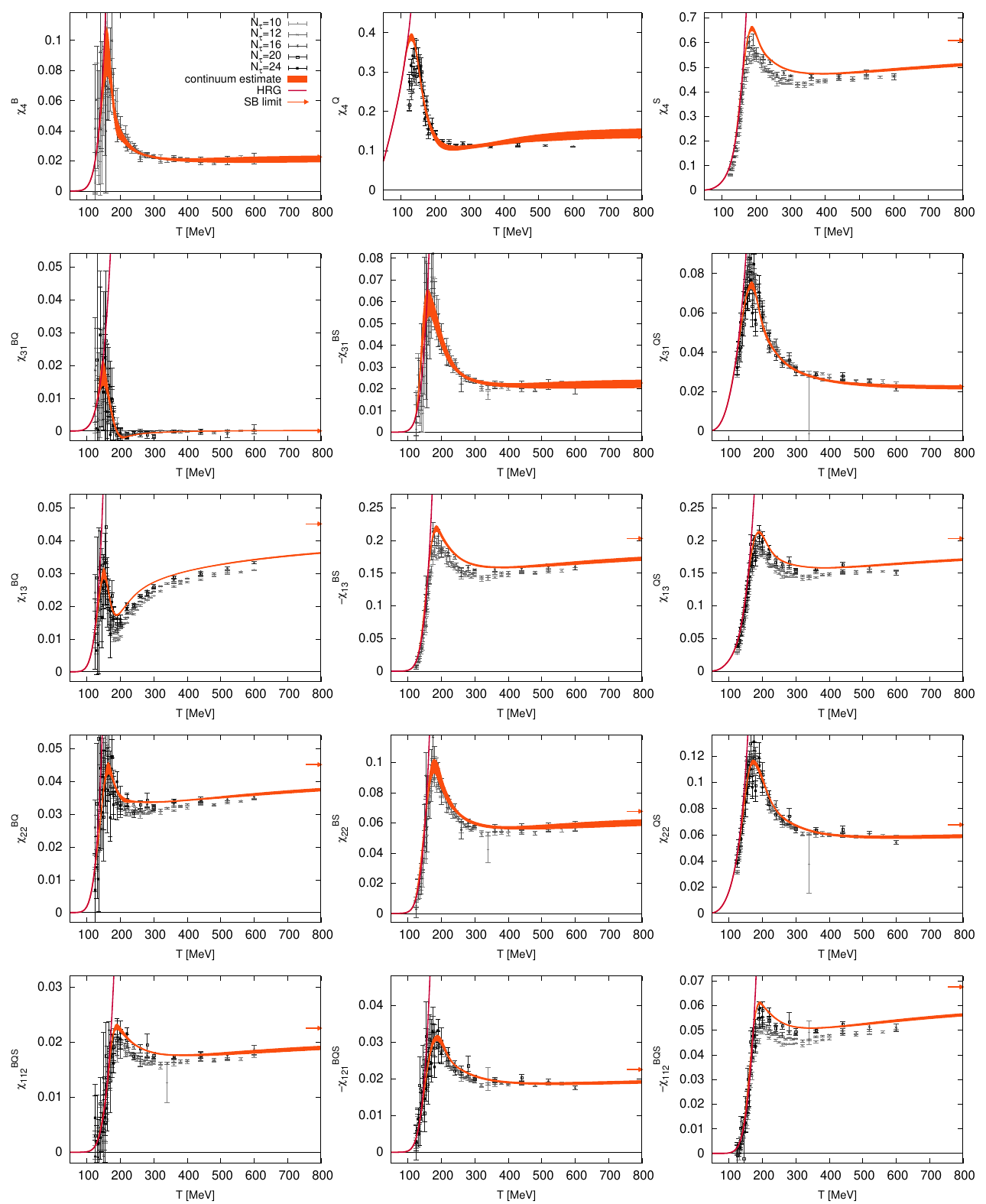}
        \caption{Continuum estimate of all fourth order susceptibilities (orange). The Stefan-Boltzmann limits of the susceptibilities are indicated by the  arrows. The gray points show the finite-$N_\tau$ lattice data used to construct the estimate. The solid lines show the ideal HRG model predictions.
        } \label{fig:chi_o4}
\end{figure*}

The procedure we describe in this work is completely determined by the conserved charge susceptibilities $\chi^{BQS}_{ijk}$ of order 2 and 4. We construct continuum estimates for all these quantities, based on lattice results with $N_\tau = 10,12,16,20,24$ timeslices, as well as on HRG model results \cite{Alba:2017mqu}. The lattice results are obtained with the 4stout action with physical quark masses, details on which can be found in Ref.~\cite{Bellwied:2015lba}. The lattices we employ have an aspect ratio $LT=4$, except for the $64^3 \times 24$ lattice.

The final susceptibilities are obtained by merging three different regimes. For temperatures below the reach of lattice simulations, namely $T \leq 120 \MeV$, we calculate all susceptibilities, as well as the $T$-derivatives of the second order ones, in the ideal HRG model. 
For the high temperature regime ($T> 200$~MeV), we perform a combined fit to lattice data in $1/N_\tau^2$ (linear) and $1/T^n$ (\mbox{polynomial} of order $n=2$, 3, 4 or 5). This way, we enforce the approach to the Stefan-Boltzmann limit at $T\rightarrow \infty$. 
Similarly, we perform a combined fit to the same lattice data in $1/N_\tau^2$ (linear) and $T$ (cubic spline) in the range $125 \MeV \leq T \leq 600 \MeV$. This way, we can ensure a proper overlap of the intermediate-$T$ results with both the low- and high-$T$ regimes.  
The three results are then merged with a subsequent spline fit, whereby the intermediate and high-$T$ regimes are overlapped to ease the merging. For the same reason, the HRG model results are included with a fictitious 5-10$\%$ error. 

Finally, the low-$T$ data from the spline fit are replaced by exact ideal HRG results, in order to avoid even very small deviations, which can combine into large differences at the level of the determination of the $\lambda_2$ coefficients.
This replacement is enforced at around \mbox{$T \sim 100$ MeV}, after ensuring that the ideal HRG results and the spline fit agree in the range \mbox{$100 < T < 120$ MeV}. 
The final result is then given directly by the HRG model at low-$T$, and from the best-fit result for larger values of the temperature. 
Note that the error introduced to the exact HRG data in the low-$T$ regime is taken to be the same relative error as the first point of the spline fit over the combined HRG, intermediate and high-$T$ lattice datasets.

In Appendix~\ref{apdx:chis_table}, we list the temperature ranges used for the high-$T$ and intermediate-$T$ fits to lattice data, the order of the polynomial fits, the lattice sizes used and the temperatures below which exact HRG results are included, for each of the susceptibilities.

In Figs.~\ref{fig:chi_o2},~\ref{fig:chi_o4} we show the continuum estimated results for all susceptibilities of order 2 and 4. The $T$-derivatives of order 2 susceptibilities are also shown in Fig.~\ref{fig:chi_o2}. 
In all plots, the lattice results at finite $N_\tau$ included in the fits are shown in shades of grey, and the value of the SB limit is indicated by an arrow on the right side of the plot. Although we show all results up to a temperature $T = 800 \MeV$, by enforcing the correct Stefan-Boltzmann limit behavior we ensure that, by construction, all susceptibilities are available also above this value.

These susceptibilities are publicly accessible through~\cite{jahan_2025_15123623}.

\section{\label{sec:results} Results\protect\\}

\subsection{\label{}Expansion coefficients and limits of applicability}

Once all susceptibilities of order 2 and 4 are fixed, our construction is fully determined. The next step is to construct the coefficient $\lambda_2(T)$ in Eq.~\eqref{eq:generalizedlambda}.

\begin{figure}[!h]
    \includegraphics[width=\linewidth]{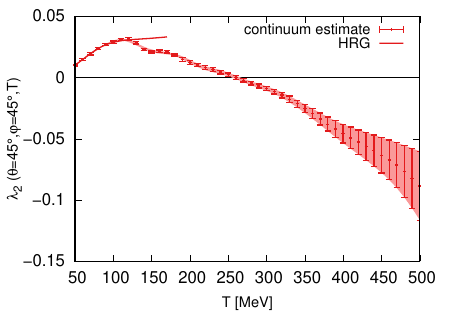}
    \includegraphics[width=\linewidth]{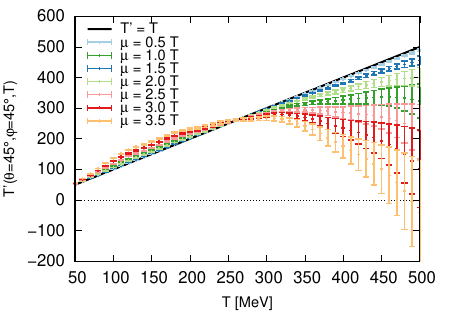}
    \caption{Top: temperature dependence of the $\lambda_2$ coefficient resulting from the susceptibilities shown in Figs.~\ref{fig:chi_o2} and \ref{fig:chi_o4}, along with a slightly smoother version obtained as detailed in the text (pink band). Bottom: the resulting $T^\prime(T)$ for several values of $\hmu$ at $(\theta = 45^\circ, \varphi = 45^\circ)$.
    } 
    \label{fig:lambda2_BQS}
\end{figure}

By scanning through all possible directions in the 3D space of $(\mu_B,\mu_Q,\mu_S)$, we observe a significant change in the $T$ dependence of this coefficient. For example, in the $\mu=\mu_B$ direction considered in 
Ref.~\cite{Borsanyi:2022qlh}, $\lambda^B_2(T)$ is a positive unimodal function of $T$, which becomes consistent with 0 around $T \sim 350$ MeV, and remains so at higher temperatures.
However, in other directions $\lambda_2(T)$ turns negative, as shown in Fig.~\ref{fig:lambda2_BQS} (top panel) for the direction where $(\theta = 45^\circ, \varphi = 45^\circ)$. We choose to show this direction because it includes all three chemical potentials, and thus the results depend on all 21 susceptibilities shown in Figs.~\ref{fig:chi_o2},~\ref{fig:chi_o4}. 
In order to limit the impact of numerical effects, especially on quantities containing temperature derivatives, we perform a very light smoothing of the coefficients $\lambda_2(T)$, which we obtain by combining the results of five different spline fits. It appears clearly in Fig.~\ref{fig:lambda2_BQS} (top) that the effect is very mild.

\begin{figure}
    \includegraphics[width=\linewidth]{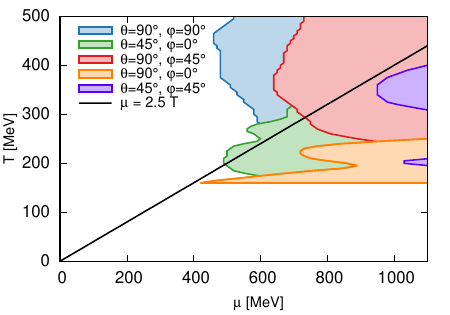}
        \caption{Regions where our expansion breaks down according to the criterion defined in the text, for different values of the angles.} \label{fig:stability}
\end{figure}

Unlike what was done in Ref. \cite{Borsanyi:2021sxv} in the $\mu = \mu_B$ direction, it is currently not possible to test the validity of the TExS in any direction $(\theta,\varphi)$, by checking up to which $\hat{\mu}$ value the expansion computed at NNLO agrees with the NLO expansion. 
This is because computing $\lambda_4(T)$ in 4D requires to have all mixed $B,Q,S$ susceptibilities of order 2, 4 and 6 (as well as some of their derivatives w.r.t. $T$), which are not known with sufficient precision to allow a meaningful analysis. 
Hence, we estimate the limits of applicability of our expansion, based on physics arguments, as follows.
The generalized density $X_1(T)$ is expected to be a monotonically increasing function for a fixed value of $\hat{\mu}$, in the absence of a first-order phase transition. As defined in Eq.~\ref{eq:4D-TExS}, $X_1(T,\hat{\mu})$ is constructed from $X_2(T',0)$, which is also a monotonically increasing function of $T'$. Thus, we conclude that $T'(T)$ must be a monotonically increasing function, and if it isn't our expansion breaks down. This behavior is purely due to the truncation of the expansion of $T'(T)$ in terms of $\hat\mu$ at LO.

We estimate the range of validity of our expansion by searching the value of $\hmu$ above which we observe non-monotonic behavior in $T'(T,\hmu)$, i.e. where $dT'/dT < 0$. In Fig.~\ref{fig:lambda2_BQS} (bottom panel) we show $T'(T,\hmu)$ in the direction $(\theta = 45^\circ, \varphi = 45^\circ)$ for different values of the chemical potential. For each temperature, there is a minimum value of $\hmu$ above which $dT^\prime/dT$ is negative: in this regime, our extrapolation breaks down.

We summarize in Fig.~\ref{fig:stability} our estimates for the upper limits of applicability of the TExS in terms of $\hmu$ obtained in this way, showing results for values of the angles $(\theta,\varphi)$ obtained by scanning the solid angle in steps of $\pi/4$. We note that the directions $(\theta=0^\circ, \, \varphi=0^\circ)$ (where $\hmu = \hmu_B$) and $(\theta=45^\circ, \, \varphi=90^\circ)$ (where $\hmu_Q=0$ and $\hmu_B = \hmu_S$) are not shown as in these directions the expansion does not break down for $\mu \leq 1200 \MeV$. 
Although in some directions the extrapolation breaks down below $\hmu \leq 2.5$, the dimensionful chemical potentials we cover are always above $\mu \simeq 400 \MeV$, and in most cases considerably larger. This means that in general, the range our extrapolation reaches is sufficient to cover the regime probed by heavy-ion collisions. 

\subsection{\label{subsec:thermo_finitemu} Thermodynamics at finite chemical potential\protect\\ }

Thermodynamic quantities at finite values of the chemical potentials are obtained in perfect analogy to Refs.~\cite{Borsanyi:2021sxv,Borsanyi:2022qlh}. From here on, we drop the indexes that indicate the dependence on the angles to make the notation lighter. 
We also elect to work with dimensionless quantities, hence all thermodynamic functions are normalized by the corresponding power of the temperature (e.g. $\hat{p} = p/T^4$, $\hat{s} = s/T^3$). 
Starting from the generalized density $X_1(T,\hmu)$ of Eq.~\eqref{eq:4D-TExS}, the pressure is calculated via the integral: 
\begin{align} 
    \label{eq:pressure}
    \hat{p}(T,\hmu) &= \hat{p} (T,\hmu=0) + \int_0^{\hmu}  d\hmu^\prime  X_1(T,\hmu^\prime) \, \, ,
\end{align}
and the other thermodynamic functions follow straightforwardly. The integration constant $\hat{p} (T,\hmu=0)$ is the pressure at vanishing chemical potential. Although a new determination with increased precision recently became available~\cite{Borsanyi:2025dyp}, in this work we employ the results from Ref.~\cite{Borsanyi:2013bia}, as they provide a coverage up to $T = 500 \MeV$, which is needed for hydrodynamic simulations of heavy-ion collisions. 

Starting from the pressure, the entropy is defined as:
\begin{align} \nonumber
    \hat{s} (T,\hmu) 
    &= 4 \hat{p} (T,\hmu) + T \left. \frac{d \hat{p}(T,\hmu)}{dT} \right|_\mu \\
    &= 4 \hat{p} (T,\hmu) + T \left. \frac{d \hat{p}}{dT} \right|_{\hmu} - \hmu X_1(T,\hmu) \, \, ,
    \label{eq:entropy}
\end{align}
where in the last line we have re-written the derivative at constant $\mu$ in terms of the derivative at constant $\hmu$, to which we have access on the lattice. 
The energy density follows as:
\begin{equation}
    \hat{\epsilon} (T,\hmu) = \hat{s} (T,\hmu) - \hat{p} (T,\hmu) + \hmu X_1 (T,\hmu) \, \, .
\end{equation}

The expressions for conserved charge densities $\hat{n}_i$ ($\hat{n}_B, \hat{n}_Q, \hat{n}_S$) are more involved, and can be written as:
\begin{equation}
    \hat{n}_i = \left. \frac{\partial \hat{p}}{\partial \hmu_i} \right|_T = \left. \frac{\partial }{\partial \hmu_i}\right|_T \int_0^{\hmu}  d\hmu^\prime  X_1(T,\hmu^\prime) \, ,
\end{equation}
where the derivative is written from the chain rule as:
\begin{equation}
    \frac{\partial }{\partial \hmu_i} = 
    \frac{\partial \hmu}{\partial \hmu_i} \frac{\partial }{\partial \hmu} + 
    \frac{\partial \theta}{\partial \hmu_i} \frac{\partial }{\partial \theta} +
    \frac{\partial \varphi}{\partial \hmu_i} \frac{\partial }{\partial \varphi}
    \; .
\end{equation}

Hence, the densities read:
\begin{align}
\label{eq:4DTExS_density}
    \hat{n}_i &= \frac{\partial \hat{p}}{\partial \hmu_i} 
    = \frac{\partial }{\partial \hmu_i} \int_0^{\hmu}  d\hmu^\prime  X_1(T,\hmu^\prime) 
    \\ \nonumber
    &= \frac{\hat{\mu}_i}{\hat{\mu}} X_1(T,\hmu) + \left( \frac{\partial \theta}{\partial \hmu_i} \frac{\partial }{\partial \theta} 
    + \frac{\partial \varphi}{\partial \hmu_i} \frac{\partial }{\partial \varphi} \right)  \int_0^{\hmu}  d\hmu^\prime X_1(T,\hmu^\prime) 
    \\ \nonumber
    &= \frac{\hat{\mu}_i}{\hat{\mu}} X_1(T,\hmu) 
    + \left( \frac{\partial \theta}{\partial \hmu_i} \int_0^{\hmu}  d\hmu^\prime \frac{\partial X_1(T,\hmu^\prime)}{\partial \theta} \right. 
    \\ \nonumber
    & \qquad \qquad \qquad \qquad \qquad \left. +\frac{\partial \varphi}{\partial \hmu_i} 
    \int_0^{\hmu}  d\hmu^\prime \frac{\partial X_1(T,\hmu^\prime)}{\partial \varphi} \right)
    \; ,
\end{align}
where the expressions for the derivatives of the angles w.r.t. the $B,Q,S$ chemical potentials and the derivatives of $X_1(T,\hmu)$ w.r.t. the angles are given in Appendices~\ref{apdx:mus_derivatives} and \ref{apdx:angle_derivatives}, respectively.
\\

\begin{figure*}
    \includegraphics[width=\textwidth]{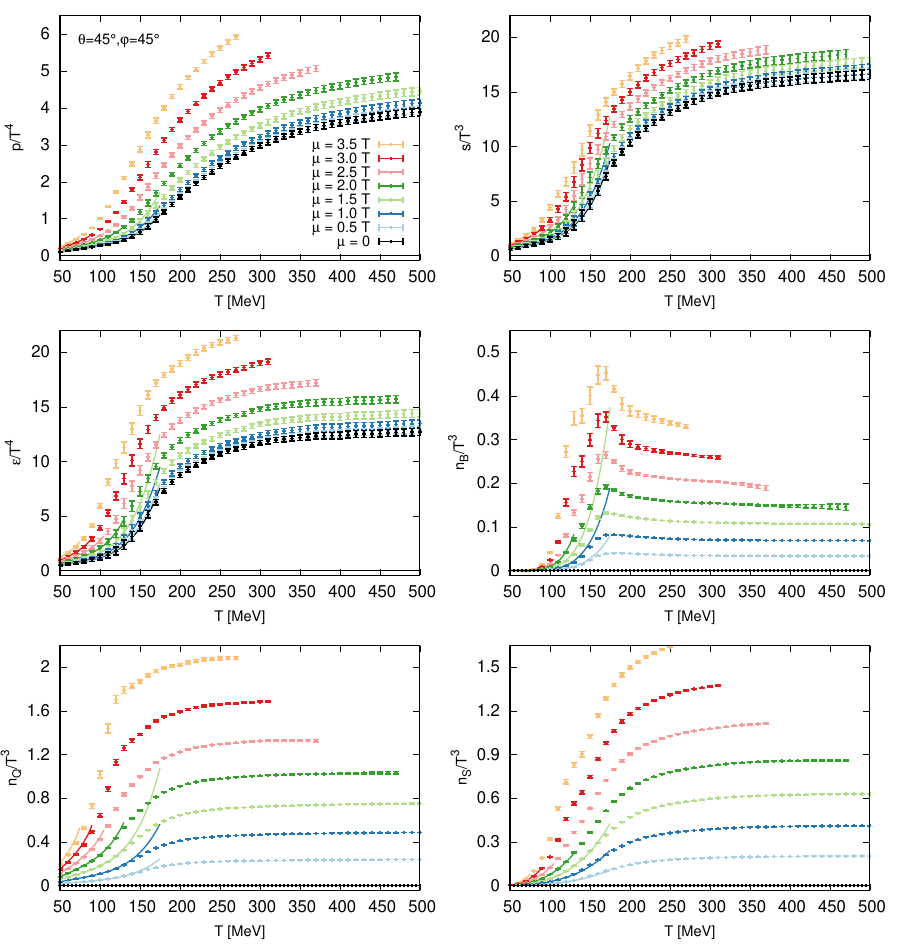}
    \caption{Pressure along with entropy, energy, baryon, electric charge and strangeness densities as functions of the temperature for increasing values of the chemical potential $\hmu$, for $(\theta = 45^\circ$, $\varphi = 45^\circ)$. The solid lines show predictions from the HRG model.
    }\label{fig:thermo}
\end{figure*}

We present in Fig.~\ref{fig:thermo}, as an example, our results for the thermodynamic quantities at different \mbox{chemical} \mbox{potentials} $\mu/T =0-3.5$ for the choice of angles $(\theta=45^\circ, \,\varphi=45^\circ)$, corresponding to a direction where $\mu_B/\sqrt{2} = \mu_Q = \mu_S$, same as in Fig.~\ref{fig:lambda2_BQS}.
We show results for pressure, entropy density, energy density, as well as baryon, electric charge and strangeness number densities. All results are compared to predictions from the ideal HRG model. 
Curves corresponding to larger values of $\mu/T$ stop at lower temperatures because we limit the values of the dimensionful chemical potentials to $\mu_B \leq 670 \MeV$, $\mu_Q \leq 475 \MeV$ and $\mu_S \leq 475 \MeV$, which nonetheless cover the range of interest for hydrodynamic simulations of heavy-ion collisions. 
These ranges are well within the range of applicability as shown in Fig.~\ref{fig:stability} for this choice of angles.

We see that the TExS produces results that do not suffer from apparent unphysical behavior, and the agreement with HRG model predictions is excellent up to temperatures around the QCD transition. As expected, deviations between the HRG results and our extrapolation appear at lower temperature when the chemical potential is increased, signaling the bending of the transition in the QCD phase diagram.

\section{\label{sec:conclusions}Conclusions\protect\\}

We presented a new method to compute the equation of state of QCD in the full 3D space of $B,Q,S$ chemical potentials, generalizing a recently proposed procedure. Our results are based on continuum estimated susceptibilities of order 2 and 4 obtained with the 4stout action on lattices with $N_\tau = 10,12,16,20,24$ timeslices. Through a change of coordinates, we constructed a procedure to carry out only 1D extrapolations by varying the direction in the chemical potential space along which they are performed.
We carried out the expansion to LO, which however corresponds to NLO in a Taylor expansion, involving all susceptibilities of order 4. Our results include the uncertainties based on the continuum estimated susceptibilities and the zero-density equation of state, although they don't include a complete systematic error analysis.

We provide the continuum estimated susceptibilities in the temperature range $T = 30 - 800 \MeV$ \cite{jahan_2025_15123623}, covering the whole regime of interest for nuclear collisions, but by construction they encode the correct high temperature limit, and can thus be trusted for higher temperatures too.

In order to estimate the limit of applicability of the TExS at LO, we performed an analysis of the effective temperature $T^\prime(T)$, and mapped the regions where the extrapolation breaks down. We found that, although limited in some specific directions in the phase diagram, our method overall provides a broader coverage in the chemical potentials compared to existing constructions based on the Taylor expansion.

As an example, we showed our results for the pressure, entropy density, energy density, as well as baryon number, electric charge and strangeness densities, for chemical potentials up to $\hmu = 3.5$ in the direction defined by $(\theta = 45^\circ, \varphi=45^\circ)$. We found that no unphysical behavior is present in the range we cover, and predictions from the HRG model are in excellent agreement with our expansion.

The code developed to obtain this equation of state will be included in a forthcoming update of the MUSES calculation engine \cite{MUSES_WEBSITE, muses_calculation_engine_2025_v1, ReinkePelicer:2025vuh}, to provide an open-access platform to compute the equation of state of QCD while independently varying all chemical potentials, in the temperature range $T = 30 -500 \MeV$.

As our procedure is based upon a rigorous expansion scheme, it is systematically improvable by including higher orders in the susceptibilities pool. Should susceptibilities of order 6 become available in the future, they can be readily incorporated, which will likely result in an even wider coverage of the phase diagram.


\begin{acknowledgements}
This material is based upon work supported by the National Science Foundation under grants No. PHY-2208724 and PHY-2116686, and within the framework of the MUSES collaboration, under Grant  No. OAC-2103680. This material is also based upon work supported by the U.S. Department of Energy, Office of Science, Office of Nuclear Physics, under Awards Number DE-SC0022023 and Number DE-SC0025025, as well as by the National Aeronautics and Space Agency (NASA) under Award Number 80NSSC24K0767.
This work is also supported by the state of North Rhine-Westphalia, Germany (MKW NRW) under the funding code NW21-024-A.
An award for computer time was provided by the U.S. Department of Energy’s (DOE) Innovative and Novel Computational Impact on Theory and Experiment (INCITE) Program. This research used resources from the Argonne Leadership Computing Facility, a U.S. DOE Office of Science user facility at Argonne National Laboratory, which is supported by the Office of Science of the U.S. DOE under Contract No. DE-AC02-06CH11357. The authors gratefully acknowledge the Gauss Centre for Supercomputing e.V. (\url{www.gauss-centre.eu}) for funding this project by providing computing time on the GCS Supercomputers Juwels-Booster at Juelich Supercomputer Centre and HAWK at H\"ochstleistungsrechenzentrum Stuttgart.
This work is also supported by the Hungarian National Research, Development and Innovation Office under Project No. FK 147164 and Grant No. NKKP Excellence 151482.
\end{acknowledgements}

\appendix

\section{\label{apdx:chis_table}Construction of the continuum estimated susceptibilities}

We list in Table~\ref{tab:chis} the temperature ranges used for the low-$T$ and high-$T$ fits to lattice data, the order of the polynomial fits for the high-$T$ part, the lattice sizes used and the temperatures up to which exact HRG results are included, for each of the susceptibilities.
The second column corresponds to the temperature range from HRG data used in the final fit for each quantity.

\begin{table*}[h]
    \centering
    \begin{tabular}{|c|c|c|c|c|c|c|}
        \hline
        \textbf{}
        & \multicolumn{3}{|c|}{\textbf{$T$ range [MeV]}} 
        & \textbf{Order of} 
        & \textbf{$N_\tau$ slices} 
        & \textbf{Merging $T$ with} 
        \\ 
        \cline{2-4}
        \textbf{} 
        & \textbf{HRG} 
        & \textbf{LQCD (spline)} 
        & \textbf{LQCD ($1/T^n$)} 
        & \textbf{$1/T^n$ fit}
        & \textbf{used} 
        & \textbf{HRG data [MeV]} 
        \\ \hline
        $\chi^{B}_{2}$       & $30-120$ & $120-300$ & $200-600$ & $n=4$ & 10, 12, 16 & 107 \\ \hline 
        $d\chi^{B}_{2}/dT$   & $30-120$ & $120-300$ & $200-600$ & $n=4$ & 10, 12, 16 & 98 \\ \hline 
        $\chi^{Q}_{2}$       & $30-120$ & $120-220$ & $210-600$ & $n=4$ & 16, 20, 24 & 100 \\ \hline 
        $d\chi^{Q}_{2}/dT$   & $30-120$ & $120-210$ & $210-600$ & $n=4$ & 16, 20, 24 & 112 \\ \hline 
        $\chi^{S}_{2}$       & $30-120$ & $130-300$ & $220-600$ & $n=5$ & 12, 16, 20 & 123 \\ \hline 
        $d\chi^{S}_{2}/dT$   & $30-120$ & $130-300$ & $220-600$ & $n=5$ & 12, 16, 20 & 115 \\ \hline 
        $\chi^{BQ}_{11}$     & $30-120$ & $120-210$ & $210-600$ & $n=4$ & 10, 12, 16 & 92 \\ \hline 
        $d\chi^{BQ}_{11}/dT$ & $30-120$ & $120-210$ & $210-600$ & $n=4$ & 10, 12, 16 & 92 \\ \hline 
        $\chi^{BS}_{11}$     & $30-120$ & $120-210$ & $210-600$ & $n=4$ & 10, 12, 16 & 113 \\ \hline 
        $d\chi^{BS}_{11}/dT$ & $30-120$ & $120-210$ & $210-600$ & $n=4$ & 10, 12, 16 & 107 \\ \hline 
        $\chi^{QS}_{11}$     & $30-120$ & $120-210$ & $210-600$ & $n=4$ & 12, 16, 20 & 108 \\ \hline 
        $d\chi^{QS}_{11}/dT$ & $30-120$ & $120-210$ & $210-600$ & $n=4$ & 12, 16, 20 & 107 \\ \hline 
        $\chi^{B}_{4}$       & $30-120$ & $120-300$ & $200-600$ & $n=3$ & 10, 12, 16 & 116 \\ \hline 
        $\chi^{Q}_{4}$       & $30-120$ & $120-300$ & $200-600$ & $n=3$ & 16, 20, 24 & 105 \\ \hline 
        $\chi^{S}_{4}$       & $30-120$ & $120-300$ & $200-600$ & $n=3$ & 10, 12, 16 & 136 \\ \hline 
        $\chi^{BQ}_{22}$     & $30-120$ & $120-240$ & $240-600$ & $n=3$ & 10, 12, 16, 20, 24 & 90 \\ \hline 
        $\chi^{BS}_{22}$     & $30-120$ & $125-300$ & $200-600$ & $n=3$ & 10, 12, 16 & 93 \\ \hline 
        $\chi^{QS}_{22}$     & $30-120$ & $120-300$ & $230-600$ & $n=3$ & 10, 12, 16, 20, 24 & 90 \\ \hline 
        $\chi^{BQ}_{13}$     & $30-120$ & $120-300$ & $220-600$ & $n=3$ & 10, 12, 16, 20, 24 & 86 \\ \hline 
        $\chi^{BS}_{13}$     & $30-120$ & $120-300$ & $230-600$ & $n=3$ & 10, 12, 16 & 83 \\ \hline 
        $\chi^{QS}_{13}$     & $30-120$ & $120-340$ & $290-600$ & $n=3$ & 10, 12, 16, 20, 24 & 74 \\ \hline 
        $\chi^{BQ}_{31}$     & $30-120$ & $130-300$ & $245-600$ & $n=2$ & 10, 12, 16, 20, 24 & 110 \\ \hline 
        $\chi^{BS}_{31}$     & $30-120$ & $120-400$ & $230-600$ & $n=3$ & 10, 12, 16 & 84 \\ \hline 
        $\chi^{QS}_{31}$     & $30-120$ & $120-300$ & $200-600$ & $n=3$ & 10, 12, 16, 20, 24 & 91 \\ \hline 
        $\chi^{BQS}_{211}$   & $50-130$ & $145-300$ & $200-600$ & $n=3$ & 10, 12, 16, 20, 24 & 95 \\ \hline 
        $\chi^{BQS}_{121}$   & $30-120$ & $120-250$ & $220-600$ & $n=3$ & 10, 12, 16, 20, 24 & 83 \\ \hline 
        $\chi^{BQS}_{112}$   & $30-120$ & $120-360$ & $220-600$ & $n=3$ & 10, 12, 16, 20, 24 & 94 \\ \hline 
    \end{tabular}
    \caption{Summary of the temperature ranges used in our fits, the order of the $1/T^n$ polynomial fits used for the high-$T$ lattice data, the timeslices used for lattice continuum estimates, and the temperature up to which actual HRG data is used for each susceptibility employed in our construction.}
    \label{tab:chis}
\end{table*}

\section{\label{apdx:mus_derivatives}Derivatives of \texorpdfstring{$\theta,\varphi$}{} angles with respect to \texorpdfstring{$B,Q,S$ chemical potentials}{}}

In order to compute the densities for conserved charges $B,\,Q$ and $S$ defined in Eq.~\ref{eq:4DTExS_density}, we derive the two spherical angles $\theta, \varphi$ w.r.t. $\hmu_B$, $\hmu_Q$ and $\hmu_S$, using the definitions from Eq.~\eqref{eq:anglestomus}:
\begin{align}
    \frac{\partial \theta}{\partial \hat{\mu}_B} 
    &= -\frac{\sqrt{\hat{\mu}_Q^2 + \hat{\mu}_S^2}}{\hat{\mu}^2} = -\frac{s_\theta}{\hat{\mu}} \; ,
    \\ \nonumber
    \frac{\partial \theta}{\partial \hat{\mu}_Q}
    &= \frac{\hat{\mu}_B\hat{\mu}_Q}{\hat{\mu}^2 \sqrt{\hat{\mu}_Q^2 + \hat{\mu}_S^2}} = \frac{c_\theta c_\varphi}{\hat{\mu}} \; ,
    \\ \nonumber
    \frac{\partial \theta}{\partial \hat{\mu}_S} 
    &= \frac{\hat{\mu}_B\hat{\mu}_S}{\hat{\mu}^2 \sqrt{\hat{\mu}_Q^2 + \hat{\mu}_S^2}} = \frac{c_\theta s_\varphi}{\hat{\mu}} \; ,
    \\ \nonumber
    \frac{\partial \varphi}{\partial \hat{\mu}_B} &= 0 \; ,
    \\ \nonumber
    \frac{\partial \varphi}{\partial \hat{\mu}_Q} &= -\frac{\hat{\mu}_S}{\hat{\mu}_Q^2 + \hat{\mu}_S^2} = -\frac{s_\varphi}{\hat{\mu}s_\theta} \; ,
    \\ \nonumber
    \frac{\partial \varphi}{\partial \hat{\mu}_S} &=  \frac{\hat{\mu}_Q}{\hat{\mu}_Q^2 + \hat{\mu}_S^2} = \frac{c_\varphi}{\hat{\mu}s_\theta} \;.
\end{align}

\section{\label{apdx:angle_derivatives}Derivatives of generalized susceptibilities and \texorpdfstring{$T^\prime$}{} with respect to \texorpdfstring{$\theta,\varphi$}{} angles}

We detail here the calculation of the angle derivative of the generalized charge density $X_1(T,\hmu)$, needed in Eq.~\eqref{eq:4DTExS_density} to compute the $B,\,Q,\,S$ densities, using Eq.~\eqref{eq:4D-TExS}:
\begin{align}
    \frac{\partial X_1(T,\hmu^\prime)}{\partial \Omega} 
    &=  \frac{\partial}{\partial \Omega} \left[ \frac{\overline{X}_1( \hat{\mu})}{\overline{X}_2} X_2(T'(T,\hat{\mu}), 0) \right]  
    \\ \nonumber
    & \mkern-18mu = \frac{\partial}{\partial \Omega} \left[ \frac{\overline{X}_1( \hat{\mu})}{\overline{X}_2} \right] X_2(T',0)
    \\ \nonumber
    & \quad + \frac{\overline{X}_1( \hat{\mu})}{\overline{X}_2} \; \partial_\Omega  X_2(T', 0) \\ \nonumber
    & \mkern-18mu = \frac{\overline{X}_2 \, \partial_\Omega \overline{X}_1(\hat{\mu}) - \overline{X}_1(\hat{\mu}) \, \partial_\Omega \overline{X}_2}{{\overline{X}_2}^2} X_2(T', 0)   
    \\ \nonumber
    & \quad + \frac{\overline{X}_1( \hat{\mu})}{\overline{X}_2(0)} X_2'(T',0) \; \partial_\Omega T'(T,\hat{\mu})
    \, ,
\end{align}
and the derivatives of $T^\prime$ w.r.t. angles read:
\begin{align}
\label{eq:dTprimedangle}
    \partial_\Omega T'(T,\hat{\mu}) 
    &= T \,\hat{\mu}^2 \, \partial_\Omega \lambda_2  \\ \nonumber
    & \mkern-70mu = \frac{\hat{\mu}^2}{6 {\overline{X}_2}^2\left(X'_2\right)^2}\Bigg[
    \overline{X}_2 \left( X_2 \overline{X}_4 - \overline{X}_2 X_4 \right) 
    \partial_\Omega X_2^{\prime} \\ \nonumber
    & \qquad\quad + X_2^{\prime} \bigg(
    \partial_\Omega \overline{X}_2 \, X_2 \, \overline{X}_4
    + {\overline{X}_2}^2 \, \partial_\Omega X_4  \\ \nonumber
    & \qquad\qquad\qquad -\!\overline{X}_2 \Bigl( \partial_\Omega \overline{X}_4 \, X_2 + X_4 \, \partial_\Omega X_2 \Bigl)
    \bigg) \Bigg]
\end{align}
with $\partial_\Omega \equiv \left.\frac{\partial}{\partial\Omega}\right|_{T,\hat{\mu}}$ used for convenience, where $\Omega ={\theta, \varphi}$.
The derivatives of $X_2$, $X_4$, of their SB limits and of $\overline{X}_1(\hat{\mu})$ w.r.t. $\theta$ and $\varphi$ follow straightforwardly from 
Eqs.~\eqref{eq:X1}, ~\eqref{eq:X2} and \eqref{eq:X4}.
\\

At last, we derive the expressions for the derivatives of $X_{2/4}(T)$ w.r.t. the angles $\theta, \varphi$ entering Eq.~\eqref{eq:dTprimedangle}, based on Eqs.~\eqref{eq:X2} and \eqref{eq:X4}:
\begin{align}
    \frac{\partial X_2}{\partial \theta} & = 
    - 2 c_{\theta} s_\theta \; \chi_2^B 
    + 2 c_{\theta} s_\theta c_{\varphi}^2 \; \chi_2^Q
    + 2 c_\theta s_{\theta} s_{\varphi}^2 \; \chi_2^S 
    \\ \nonumber & \quad 
    + 2\bigl({c_\theta}^2 - {s_\theta}^2 \bigl) c_{\varphi} \; \chi_{11}^{BQ}
    + 2\bigl({c_\theta}^2 - {s_\theta}^2 \bigl) s_{\varphi} \;\chi_{11}^{BS}
    \\ \nonumber & \quad 
    + 4 c_\theta s_{\theta}c_{\varphi}s_{\varphi} \; \chi_{11}^{QS} \; ,
    \\ \nonumber \\
    \frac{\partial X_2}{\partial \varphi} & = 
    - 2 {s_\theta}^2 c_{\varphi} s_\varphi \; \chi_2^Q
    + 2 {s_\theta}^2 c_\varphi s_{\varphi} \; \chi_2^S 
    \\ \nonumber & \quad 
    - 2 {c_\theta} {s_\theta} s_{\varphi} \; \chi_{11}^{BQ}
    + 2 {c_\theta} {s_\theta} c_{\varphi} \;\chi_{11}^{BS}
    \\ \nonumber & \quad 
    + 4 {s_\theta}^2 \bigl({c_\varphi}^2 - {s_\varphi}^2\bigl) \; \chi_{11}^{QS} \; ,
\end{align}
\begin{align}
    \frac{\partial X_4}{\partial \theta} & = 
    - 4 s_\theta^3 s_\theta \; \chi_4^B 
    + 4 c_\theta s_\theta^3 c_\varphi^4 \; \chi_4^Q
    + 4 c_\theta s_\theta^3 s_\varphi^4 \; \chi_4^S 
    \\ \nonumber & \quad 
    + 4 \bigl(3c_\theta^2 s_\theta^2 - s_\theta^4\bigl) c_\varphi^3 \; \chi_{13}^{BQ}
    + 4 \bigl(3c_\theta^2 s_\theta^2 - s_\theta^4 \bigl) s_\varphi^3 \; \chi_{13}^{BS}
    \\ \nonumber & \quad 
    + 16 c_\theta s_\theta^3 c_\varphi s_\varphi^3 \; \chi_{13}^{QS}
    + 4 \bigl(c_\theta^4 - 3c_\theta^2 s_\theta^2 \bigl) c_\varphi \; \chi_{31}^{BQ}
    \\ \nonumber & \quad 
    + 4 \bigl(c_\theta^4 - 3c_\theta^2 s_\theta^2 \bigl) s_\varphi \; \chi_{31}^{BS}
    + 16 c_\theta s_\theta^3 c_\varphi^3 s_\varphi \; \chi_{31}^{QS}
    \\ \nonumber & \quad 
    + 12 \bigl(c_\theta^3 s_\theta - c_\theta s_\theta^3 \bigl) c_\varphi^2 \; \chi_{22}^{BQ}
    + 24 c_\theta s_\theta^3 c_\varphi^2 s_\varphi^2 \; \chi_{22}^{QS}
    \\ \nonumber & \quad 
    + 12 \bigl(c_\theta^3 s_\theta - c_\theta s_\theta^3 \bigl) s_\varphi^2 \; \chi_{22}^{BS}
    \\ \nonumber & \quad 
    + 12 \bigl(3c_\theta^2 s_\theta^2 - s_\theta^4 \bigl) c_\varphi s_\varphi^2 \; \chi_{112}^{BQS}
    \\ \nonumber & \quad 
    + 12 \bigl(3c_\theta^2 s_\theta^2 - s_\theta^4 \bigl) c_\varphi^2 s_\varphi \; \chi_{121}^{BQS}
    \\ \nonumber & \quad 
    + 12 \bigl(c_\theta^3 s_\theta - c_\theta s_\theta^3 \bigl) c_\varphi s_\varphi \; \chi_{211}^{BQS} 
    \; ,
\end{align}
\begin{align}
    \frac{\partial X_4}{\partial \varphi} & = 
    - 4 s_\theta^4 c_\varphi^3 s_\varphi \; \chi_4^Q
    + 4 s_\theta^4 c_\varphi s_\varphi^3 \; \chi_4^S 
    \\ \nonumber & \quad 
    - 12 c_\theta s_\theta^3 c_\varphi^2 s_\varphi \; \chi_{13}^{BQ}
    + 12 c_\theta s_\theta^3 c_\varphi s_\varphi^2 \; \chi_{13}^{BS}
    \\ \nonumber & \quad
    + 4 s_\theta^4 \bigl(3 c_\varphi^2 s_\varphi^2 - s_\varphi^4 s_\theta^2 \bigl) \; \chi_{13}^{QS}
    \\ \nonumber & \quad 
    - 4 c_\theta^3 s_\theta s_\varphi \; \chi_{31}^{BQ}
    + 4 c_\theta^3 s_\theta c_\varphi \; \chi_{31}^{BS}
    \\ \nonumber & \quad 
    + 4 s_\theta^4 \bigl(c_\varphi^4 - 3  c_\varphi^2 s_\varphi^2\bigl) \; \chi_{31}^{QS}
    \\ \nonumber & \quad 
    - 12 c_\theta^2 s_\theta^2 c_\varphi s_\varphi \; \chi_{22}^{BQ}
    + 12 c_\theta^2 s_\theta^2 c_\varphi s_\varphi \; \chi_{22}^{BS}
    \\ \nonumber & \quad 
    + 12 s_\theta^4 \bigl(c_\varphi^3 s_\varphi - c_\varphi s_\varphi^3 \bigl) \; \chi_{22}^{QS}
    \\ \nonumber & \quad
    + 12 c_\theta s_\theta^3 \bigl(2 c_\varphi^2 s_\varphi -s_\varphi^3 \bigl) \; \chi_{112}^{BQS}
    \\ \nonumber & \quad 
    + 12 c_\theta s_\theta^3 \bigl(c_\varphi^3 - 2 c_\varphi s_\varphi^2 \bigl) \; \chi_{121}^{BQS}
    \\ \nonumber & \quad 
    + 12 c_\theta^2 s_\theta^2 \bigl(c_\varphi^2 - s_\varphi^2) \; \chi_{211}^{BQS}
    \; .
\end{align}

\bibliography{apssamp}

\end{document}